# Dilatancy in the flow and fracture of stretched colloidal suspensions

## M.I. Smith

School of Engineering, University of Edinburgh, Kings Buildings, Mayfield Rd, Edinburgh, EH9 3JL, UK

School of Physics and Astronomy, University of Nottingham, University Park, Nottingham, NG7 2RD

e-mail: mike.i.smith@nottingham.ac.uk

## R. Besseling

SUPA, School of Physics and Astronomy, University of Edinburgh, Kings Buildings, Mayfield Rd, Edinburgh EH9 3JZ

e-mail: rbesseli@ph.ed.ac.uk

## M.E. Cates

SUPA, School of Physics and Astronomy, University of Edinburgh, Kings Buildings, Mayfield Rd, Edinburgh, EH9 3JZ, UK

e-mail: m.e.cates@ ed.ac.uk

## V. Bertola*

School of Engineering, University of Edinburgh, Kings Buildings, Mayfield Rd, Edinburgh, EH9 3JL, UK

e-mail: v.bertola@ed.ac.uk

contacting author*

# Dilatancy in the flow and fracture of stretched colloidal suspensions.

**Concentrated particulate suspensions, commonplace in the pharmaceutical, cosmetic and food industries, display intriguing rheology. In particular, the dramatic increase in viscosity with strain rate (shear thickening and jamming) which is often observed at high volume fractions, is of strong practical and fundamental importance. Yet manufacture of these products and their subsequent dispensing often involves flow geometries substantially different from that of simple shear flow experiments. Here we show that the elongation and breakage of a filament of a colloidal fluid under tensile loading is closely related to the jamming transition seen in its shear rheology. However, the modified flow geometry reveals important additional effects. Using a model system with nearly hard-core interactions, we provide evidence of surprisingly strong viscoelasticity in such a colloidal fluid under tension. With high speed photography we also directly observe dilatancy and granulation effects, which lead to fracture above a critical elongation rate.**

When the volume fraction ($\phi$) of a colloidal suspension is increased beyond about 50%, its low-shear viscosity increases markedly, and beyond a critical shear stress the system often displays a shear-thickening or jamming transition[1-3]. Although the details of this behaviour depend on the nature of the interactions between particles, hard-sphere colloid suspensions represent an important limiting case in which the geometry rather than the energetics of particle contacts controls the behaviour. Whilst jamming transitions in such suspensions have been well investigated in simple shear flows, the same phenomena may be equally relevant in other flow geometries[4-6].

In the physics of dilatancy[7], the volume of a collection of particles must increase upon shearing to enable flow. This has been suggested as a possible mechanism for jamming in concentrated colloidal suspensions[6]. Dilation within a fixed volume of suspending liquid involves the formation of force transmitting 'clusters', whose growth eventually causes particles to encounter the air-liquid interface. This generates large capillary forces at the free surface, which can then balance the normal inter-particle forces and resist further motion. The colloidal particles can thus form spanning clusters in close contact, jamming the sample. This may then fracture into millimetre-scale 'granules'[6]. Colloidal granules can be jammed indefinitely by their capillary forces, even though the same volume of fluid and particles can equally well exist as a flowable droplet[6].

Jamming of colloids is also seen in pipe and channel flows[4,5]; here free surfaces are not present. In an extensional rheometer however[8,9] a fluid sample initially forms a cylindrical column surrounded by air, bridging two parallel, horizontal plates. As the upper plate is retracted, a filament forms which narrows and eventually breaks. An elongational flow, in contrast to more conventional shear and pipe geometries, therefore implies an increase of the interfacial area during flow. Although purely elongational flow can be achieved by exponential plate separation, a constant separation speed (as used here) is closer to fibre-spinning and other industrial processes. In these, a purely tensile loading evokes a mixed flow combining elongation and shear.

Recent studies[10,11] have demonstrated that extensional rheometry can successfully be performed on colloidal suspensions. Bischoff White et al[10] measured the tensile stresses of a ($\phi$ ~0.355) cornstarch solution. They observed a flowable filament at low extension rates but beyond this found a transition to brittle fracture. The interactions in this system are poorly characterized but clearly attractive (see Fig. 1 of Ref 10), presumably due to strong van der Waals forces. Such interactions could strongly influence the flow behaviour as they do in strongly aggregating colloids at lower densities[10]. The fundamental physics of colloidal jamming is therefore better addressed in model systems of nearly hard-sphere particles[3-6] which are forced into contact by external stresses without the added complexities of ill defined interactions.

Here we present an experimental study of the stretching of concentrated suspensions of nearly hard-sphere (PMMA[4,5]) particles. We observe that on increasing strain rate, the fluid column undergoes a transition from a liquid-like breakup to a fracture mode that we link to the dilatancy and jamming of the suspension. We also observe strong evidence of viscoelasticity in the liquid state close to this transition.

**Results**

*Strain rate dependent jamming transition*

Figure 1 shows stretched samples of a suspension with ϕ ~ 0.603. At low strain rates, a conventional filament is formed, which thins until it eventually breaks. In contrast, at high strain rates we notice angular features, with the fluid column apparently composed of 'granular' macroscopic lumps. Abrupt fracture of this column is then observed followed by separation of the two parts. After this, their angular features slowly relax, creating a pair of hemispherical droplets. For intermediate strain rates, we obtain a surprisingly complex interplay of these two types of behaviour. Initially the fluid column behaves as though composed of granules, but before complete fracture occurs, a flowable filament is recovered. Supplementary movie 1 shows this for a suspension with φ ~ 0.603 at an extension rate of 1.3 s$^{-1}$.

These results show that a colloidal suspension with nearly hard-core particle interactions can exhibit jamming and brittle fracture under tensile loading of a fluid column. To investigate this jamming transition further, we studied its dependence on volume fraction. Figure 2 shows that, when φ is altered by only a few percent, the critical strain rate shifts by 2 or 3 orders of magnitude. The insets illustrate the morphologies in the transitional regime at different volume fractions. One might expect equivalent behaviour at similar points relative to the critical strain rate. However, qualitative differences are apparent: we see increasing asymmetry at high volume fraction φ (when suspensions often pull towards one edge of the rheometer), and also increasing lumpiness. The direction of symmetry breaking varies unsystematically from experiment to experiment (unrelated to loading and/or surfaces). The lack of symmetry becomes noticeable as the granules get sufficiently large compared to the flow geometry, thus making the fluid structure non-uniform. These changes at high φ correlate with the onset of remarkable viscoelastic effects (recoil) to which we return below.

*The shear contribution to thickening*

To understand better our results for jamming of repulsive colloids under tensile stress, we recall that with linear plate separation the flow of our suspensions has a significant shear component, whose importance may depend on the sample geometry, extension rate, and volume fraction. In pure shear flow, dense hard sphere suspensions typically exhibit marked thickening within the studied range of volume fractions, and indeed, the φ dependence of the tensile jamming transition in figure 2 is very similar to that of the shear thickening transition reported in Frith et al[1]. We therefore performed traditional shear rheology on our own samples (figure 1). For φ=0.603 the suspension is strongly shear thinning at shear rates < 2 s$^{-1}$, whilst higher shear rates produce strong thickening. (Our results do not rule out a weak yield stress at very low strain rates, which is not relevant to the experiments reported here.) Comparing the critical rate of ~ 2 s$^{-1}$, to the results for extension (figure 2), we find good agreement. This supports the idea that the jamming transition in a colloidal fluid under tensile load, if not directly controlled by shear thickening, at least involves closely related physics.

Shear thickening in colloids can be hydrodynamically mediated, continuous and reversible[3], but more typical for high volume fractions is the rather sharp onset shown in figure 2. This sudden thickening has been interpreted as stress-induced jamming[6], where an applied stress that might otherwise promote a flow instead frustrates dilatancy and thus counter intuitively prevents flow.

*Role of dilatancy in jamming*

To test the contribution of dilatancy to our jammed state, we imaged the fluid surface during stretching (figure 3, supplementary movie 2). Dilatancy causes individual colloids to deform the fluid-air interface, which hence should undergo a macroscopic change from glossy to matt appearance[6]. We observe just such a change, concurrent with the onset of macroscopic lumpiness (granulation), in all our samples. This confirms the direct role of dilatancy in the granulation and subsequent fracture of colloidal suspensions under tensile loading. Whilst dilatancy has been observed in various shear-flow geometries,[5,6,12,13] its effects when stretching a fluid column appear particularly pronounced.

In bulk suspensions, dilatancy at fixed volume fraction causes an increase in the nonequilibrium osmotic pressure (particle pressure), varying roughly as $\Pi \approx \eta \dot{\varepsilon}/[1-(\phi/\phi_0)]^2$ with $\eta$ the viscosity and $\dot{\varepsilon}$ the strain rate[14]. A simple quantitative estimate of the onset of granulation equates this to the Laplace pressure $\sigma/D$ where $\sigma$ is the surface tension and $D$ the column diameter. With

σ = 0.03 N/m and η = 0.004 Pa s, this gives a critical strain rate of order $10 s^{-1}$ at ϕ =0.6, and also accounts for the decreasing trend with ϕ seen in Fig. 2.

*Elastic recoil of self-filtering filament*

Having noted the role of dilatancy in the jammed state, we now return to describe a remarkable feature of the concentrated 'liquid-like' state. When a colloidal filament is stretched at a rate below the critical strain rate, it slowly narrows as fluid drains towards the end plates. However, at ϕ ≥ 0.603 and when the filament is less than 100µm across, we see evidence of strong viscoelasticity. The filament breaks in a superficially fluid-like manner, but then recoils (figure 4). This recoil is only observed for large ϕ at strain rates close to the transition. For ϕ ~ 0.603 & 0.612, there is an initial rapid recoil which quickly slows. The full recoil takes ~ 1 & 2.1s respectively, comparable to the respective relaxation times (figure 2). The presence of such a recoil indicates that our 'fluid-like' filament can in fact support a strong tensile stress – and does so elastically.

Analysis of the change in diameter (D) of the filament prior to breakage shows an exponential decrease with time, reminiscent of a conventional viscoelastic material such as an Oldroyd-B fluid[15]. However, figure 4 shows that on final approach to breakage and recoil there is a significant departure from this behaviour, with the filament diameter remaining almost constant in time. This is evidence for the development of enhanced resistance to flow of fluid out of the thinning neck[13] which we take to indicate the onset of jamming.

In the experimental geometry the filament undergoes not only tensile stretching from the moving plates but drainage caused by surface tension[15]. This raises the possibility that jamming just prior to breakage in the fluid-like samples is caused by self-filtration[5], i.e. differential motion between particles and fluid in the thinning filament so that the particles get left in place as the fluid drains away. As ϕ increases, the particles will form a jammed bed of fixed diameter at the centre of the column. Further drainage leaves insufficient fluid to cover these particles. Stretching of this filament now requires work to be done against the capillary forces, creating a large tensile stress. This work is recoverable (elastic): when the filament breaks it recoils, releasing the stored interfacial energy. This picture implies subtle changes to the filament geometry, after jamming but prior to breakage, which cannot be fully resolved by our experiments. Moreover, a very similar sequence of events would be expected even if jamming were caused instead by conventional dilation, in response to a gradual increase of local flow rate during the experiment. However, such an increase would be at odds with our observations in the transitional regime where samples start jammed but then become fluid.

Two simple quantitative estimates lend support to a self-filtration mechanism for the eventual jamming of very thin fluid-like filaments. Firstly, jamming occurs when the filament diameter is of order 200 times the particle radius; this is comparable to the onset of self-filtration in other geometries[4,5]. Second, a radial pressure gradient estimated $\sigma/D^2$, from the laplace force , ,would – for a fluid cylinder not pinned to the particles – drive a radial flow velocity u ~ $(\sigma/D^2)$ $ka^2/\eta$, where K = $ka^2$ is the permeability of the packed bed; for $\phi$ ~ 0.6 the geometrical constant k($\phi$) is estimated[16] as 4× $10^{-3}$. For D~ 100 µm this gives u ~ 1 µm/s, leading to percent-scale changes in the fluid/particle volume ratio, which is enough to cause jamming, on a timescale of seconds. This argument mirrors the calculation of the radial flow velocity in a Newtonian fluid filament without particles[16], which is recovered by setting K ~ $D^2$

**<u>Discussion</u>**

In this study we have investigated the flow behaviour of concentrated suspensions of nearly hard-sphere colloids under tensile loading. We showed that above a critical extension rate the fluid jams and dilates, exhibiting dramatic granulation and subsequent fracture closely related to shear-thickening and jamming in simple shear flow. We demonstrated that near this transition even the flowable liquid displays surprising viscoelastic properties.

Simple estimates were given that account for the order of magnitude of the critical extension rate and its decreasing trend with volume fraction, and for the onset of a second, distinct, transition to jammed behaviour in very thin filaments prior to the final breakup. These estimates imply that the second transition, but not the first, is strongly dependent on the particle size. We leave experimental tests of this to future work.

Whilst our work aims primarily at elucidating the physical mechanisms governing dense colloidal suspensions subject to extensional deformations, it is directly relevant to a variety of practical applications where a sample volume of a colloidal suspension is separated from the rest, such as in dispensing nozzles. For example, the viscoelastic recoil of the filament that is observed just after breakup may cause nozzle clogging, with obvious consequences for operation. Moreover, the different breakup regimes observed in our experiments imply that under certain conditions it is not possible to determine with precision the position where detachment occurs and hence the dispensed volume. We hope our study will promote further work on the physics of colloidal suspensions in flows, more complex than simple shear, that have such significant implications for quality control.

**Methods**

*Sample Preparation*

Experiments were performed with PMMA spheres (radius r ~ 604nm, 5% polydispersity) sterically stabilised by a chemically grafted poly-12-hydroxy stearic acid[18] and suspended in octadecene, chosen because of its very low volatility. Refractive index matching is not as good as in some volatile solvents, but the residual interparticle attraction remains very small[18]. Suspensions were centrifuged at 2000 rpm for 12 hours to create a sediment. The volume fractions ϕ were produced by dilution of a stock solution, assuming $\phi_0$ =0.640. (Our ϕ values are then quoted to the accuracy of the subsequent dilution.) Before use, and after each dilution, the particles were thoroughly mixed using a vortex whirlie mixer to ensure uniformity.

*Extension measurements*

Extension measurements were made using a HAAKE CaBER 1 Extensional Rheometer. It consists of two parallel cylindrical stainless steel plates, 6mm in diameter and with an initial separation of 2mm. Suspensions were loaded into the rheometer to form a cylindrical column between the two plates, either by using a spatula ( ϕ > 0.6 ) or using a 1ml luer lock syringe (ϕ < 0.6). In the latter case, solutions were drawn out and deposited slowly to prevent changes in the final volume fraction[5]. The plates are then separated at a constant velocity. All strain rates are quoted using the macroscopic separation of the end plates $\dot{\varepsilon} = l_{MAX}/l_0 \Delta t$. We note that this will in general be related in a highly nonlinear way to the strain rate at different points in the sample with a decreasing strain rate with increasing separation of the plates.

The stretching of the colloidal suspensions was observed using a high speed camera (MC1311, Mikotron) equipped with a Sony 18-108 mm/f2.5 zoom lens and a 30mm extension tube at frame rates from 50 to 500fps. Back illumination was provided by a 4W Philips white light viewed through a simple diffuser. However, for high magnification imaging of the filament surface, the LED was arranged so as to reflect back off the solution. The images were captured using a x2.5 zoom lens (Edmund Optics VZM450i).

*Shear rheology measurements*

Shear rheology measurements were performed using a cone-plate rheometer, cone diameter 4cm, cone angle 2 degrees.

**Acknowledgements:**

We thank Andrew Schofield for particle synthesis. The work was funded by the UK EPSRC (Grant nos. EP/E03173/1 and EP/E005950/1 ). MEC holds a Royal Society Research Professorship.

**Author Contributions:**

MIS performed the extensional rheology measurements, RB performed the shear rheology measurements. All authors contributed to the interpretation and discussion of results, and to the drafting of the manuscript.

**Competing Financial interests statement:**

There are no competing financial interests.

**Figure legends:**

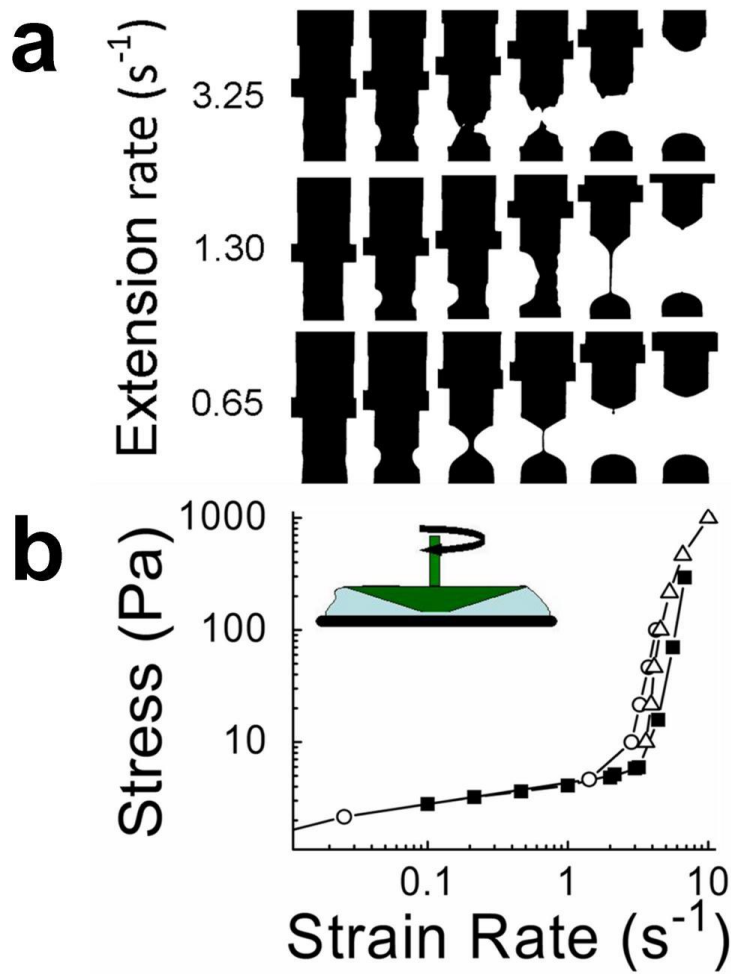

*Figure 1. Extension rate dependent morphologies of a dense colloidal suspension.* (a) At low strain rates the colloidal solution (Volume fraction ~ 0.603) forms a filament which subsequently drains towards the end plates before undergoing capillary induced break up ('liquid'). At higher rates the system jams and subsequently fractures ('jammed'). Intermediate strain rates yield an interesting composite of the two morphologies ('transition'). (b) Steady state shear rheology of the same suspension, performed using stainless steel cone-plate geometry. (■) strain rate controlled data, (△, ○) are stress controlled.

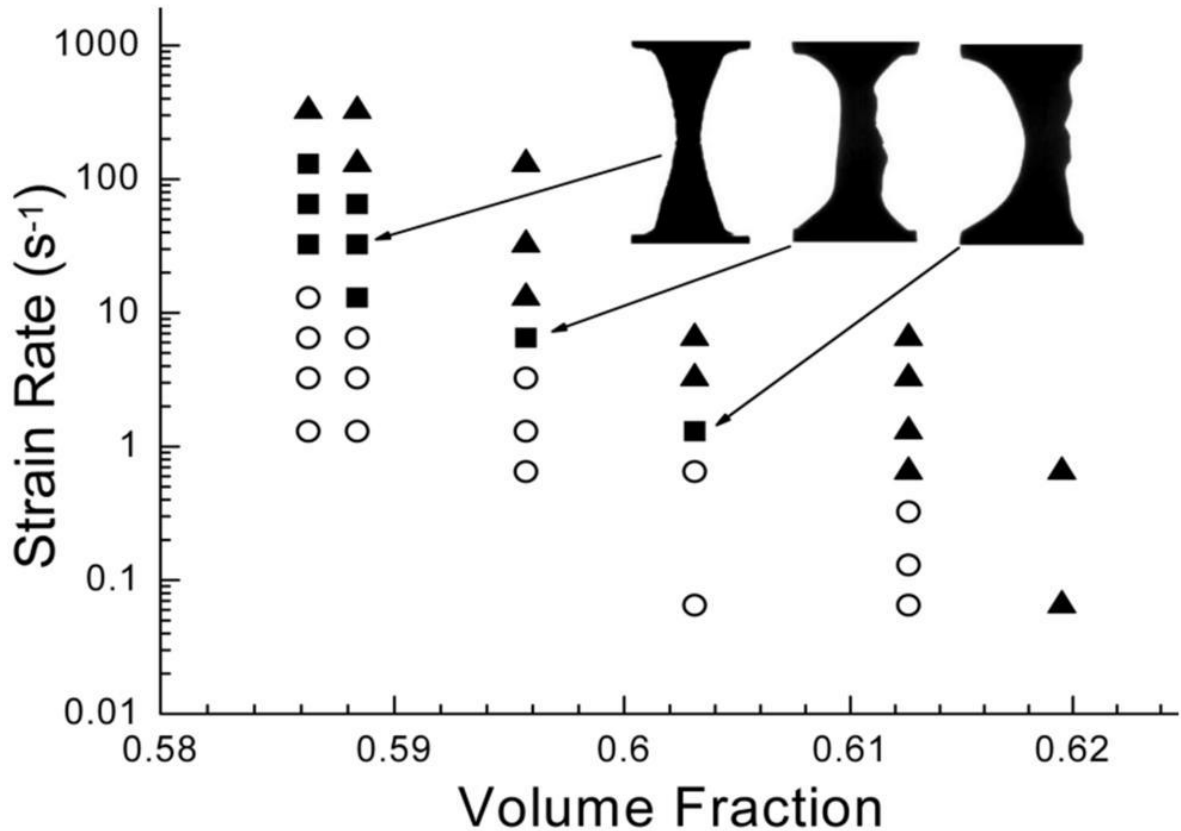

*Figure 2. Volume fraction dependence of the extensional jamming transition. 'Jammed' (▲), 'Transition' (■), 'Liquid' (○). Changes of a few percent in volume fraction result in a shift of the characteristic strain rate of the transition by orders of magnitude. The images highlight the dependence of the morphology at the transition on volume fraction (φ ~ 0.588, 0.596 & 0.603). On increasing φ, both the granule size and general asymmetry of the paste structure increases. Similar trends are observed for the jammed state.*

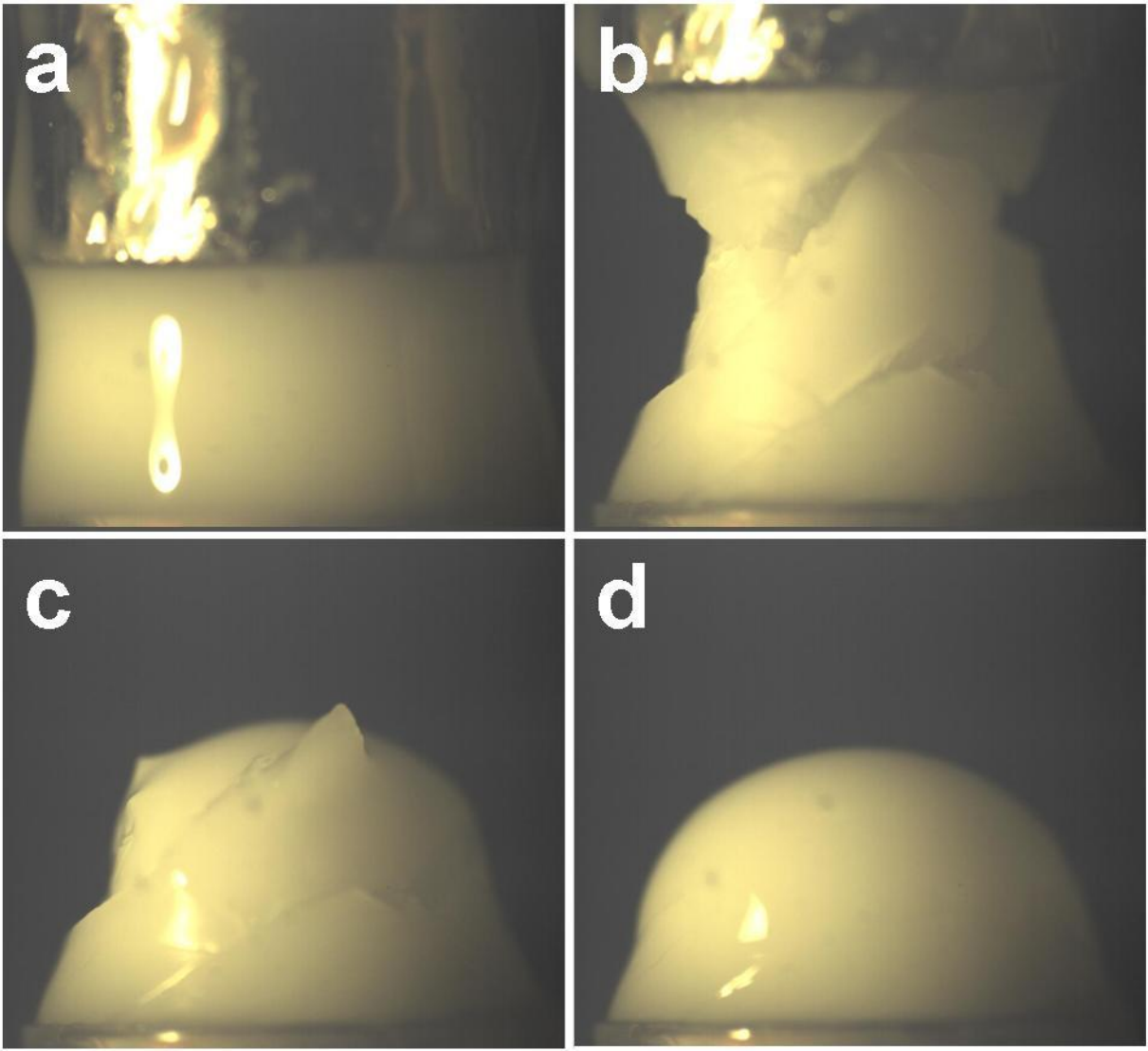

*Figure 3. Dilatancy during the jamming of colloidal suspension.* As the column of fluid is extended (Volume fraction ~ 0.603, extension rate 3.5 $s^{-1}$ ), the glossy surface seen at equilibrium (a) becomes matt as the suspending liquid retreats into the particle interstices (b). A combination of shear thickening and dilatancy lead to jamming, granulation and finally fracture. The suspension then slowly relaxes back into the equilibrium phase (c & d).

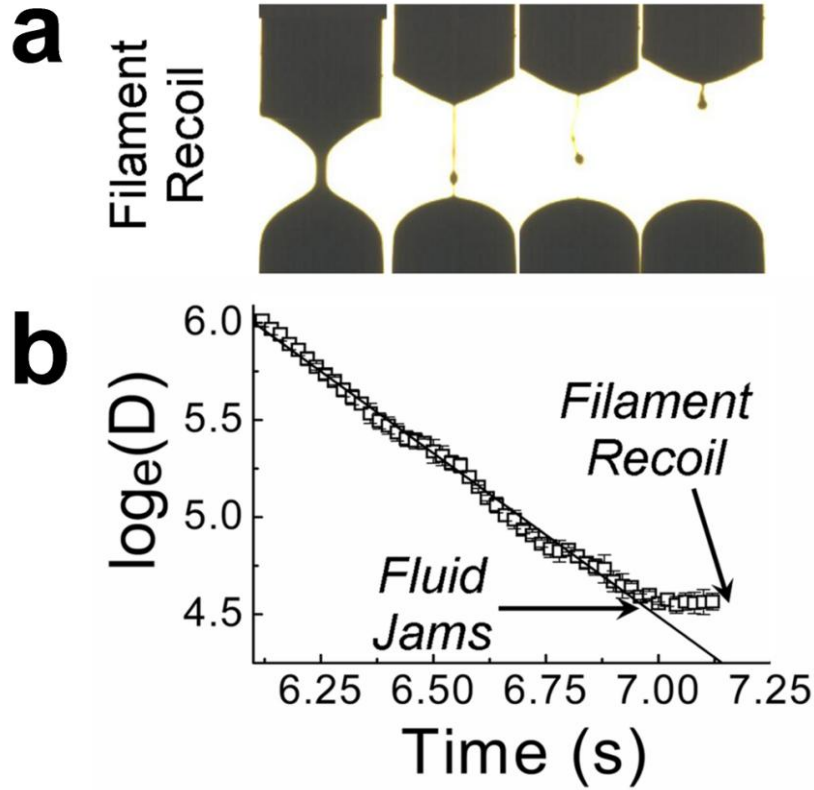

*Figure 4. Viscoelasticity of colloidal filaments. (a) Despite the apparent viscous flow characterising the thinning of the 'liquid state' filament, upon rupture we observe a strong elastic recoil of the filament towards the top plate, and a subsequent slow relaxation of this thread into the fluid state. For φ ~ 0.603 (pictured) recoil of the filament takes ~ 1s. Extension rate is $0.65s^{-1}$. (b) The natural logarithm of the mean filament diameter (measuring in 3 places) in microns as a function of time. The error bars represent the standard deviation of the filament diameter measurements. The filament thins exponentially with time as would be expected for an Oldroyd-B fluid. Departure from this indicates the development of strong additional stresses caused by jamming. As the filament continues to be stretched the diameter remains constant.*